\documentclass[twocolumn,final,epj]{svjour}
\pdfoutput=1
\usepackage[utf8]{inputenc}
\usepackage[T1]{fontenc}
\usepackage{graphicx}
\usepackage{amssymb}
\usepackage{amsmath}
\usepackage{hyperref}
\usepackage{relsize}

\newcommand\Tstrut{\rule{0pt}{1.0\normalbaselineskip}}
\newcommand\Bstrut{\rule[-0.4\normalbaselineskip]{0pt}{0pt}}

\def\eps{\varepsilon}

\DeclareMathOperator{\sgn}{sgn}

\newcommand\GR{{G_{\rm R}}}
\newcommand\Grr{{G_{rr}}}

\newcommand\Gqr{{G_{\rm qr}}}
\newcommand\Gpr{{G_{\rm pr}}}
\newcommand\Gqrd{{G_{\rm qrd}}}

\newcommand\Wqr{{W_{\rm qr}}}
\newcommand\Wpr{{W_{\rm pr}}}
\newcommand\Wrr{{W_{\rm rr}}}

\newcommand\WR{{W_{\rm R}}}

\begin{document}

\title{Linear response theory for Google matrix}

\author{
Klaus M. Frahm\inst{1} \and 
Dima L. Shepelyansky\inst{1}}

\institute{
Laboratoire de Physique Th\'eorique, IRSAMC, 
Universit\'e de Toulouse, CNRS, UPS, 31062 Toulouse, France
}

\titlerunning{Linear response theory for Google matrix}
\authorrunning{K.M.~Frahm and D.L.~Shepelyansky}

\date{\today}

\abstract{We develop the linear response theory for the Google matrix 
PageRank algorithm with respect to a general weak perturbation 
and a numerical efficient and accurate algorithm, called LIRGOMAX algorithm, 
to compute the linear response of the PageRank with respect to this 
perturbation. We illustrate its efficiency 
on the example of the English Wikipedia network with more than 5 millions 
of articles (nodes). For a group of initial nodes (or simply a pair of nodes)
this algorithm allows to identify the effective pathway between 
initial nodes thus selecting a particular subset of nodes which are most 
sensitive to the weak perturbation applied to them (injection or pumping 
at one node and absorption of probability at another node).
The further application of the reduced Google matrix algorithm (REGOMAX)
allows to determine the effective interactions between the nodes of 
this subset. General linear response theory already found
numerous applications in various areas of science
including statistical and mesoscopic physics.
Based on these grounds we argue that the developed LIRGOMAX algorithm
will find broad applications in the analysis of complex directed networks.
}


\maketitle
\section{Introduction}
Linear response theory finds a great variety of 
applications in statistical physics, stochastic processes,
electron transport,  current density correlations and dynamical systems 
(see e.g. \cite{kubo,hanggi,stone,kohn,ruelle}).
In this work we apply the approach of linear response to
Google matrices of directed networks
with the aim to characterize nontrivial interactions between nodes.

The concept of Google matrix and the related PageRank algorithm 
for the World Wide Web (WWW) has been proposed
by Brin and Page in 1998 \cite{brin}. A detailed 
description of the Google matrix construction and its properties
is given in \cite{meyer}. This approach can be applied to numerous situations 
and various directed networks \cite{rmp2015}. 

Here we develop the LInear Response algorithm
for GOogle  MAtriX (LIRGOMAX) which applies to a very general model 
of a weakly perturbed Google matrix or the related PageRank algorithm. 
As a particular application we consider a model of injection and absorption 
at a small number of nodes of the networks and test its efficiency on examples
of the English Wikipedia network of 2017 \cite{wiki2017}. However, the 
scope of LIRGOMAX algorithm is more general. Thus, for example, it can be  
also  applied to compute efficiently and accurately the PageRank sensitivity 
with respect to small modifications of individual elements of the 
Google matrix or its reduced version 
\cite{greduced,politwiki,wrwu2017,zinovyev}. 

From a physical viewpoint the approach of $\;\;\;\;$ 
injection/absorption 
corresponds to a small pumping probability at a certain network node
(or group of nodes) and absorbing probability at 
another specific node (or group of nodes).
In a certain sense such a procedure reminds lasing in random media
where a laser pumping at a certain frequency generates a response
in complex absorbing media \cite{cao}.

More specifically we select two particular nodes, one for injection and 
one for absorption, for which we use the LIRGOMAX algorithm to determine a 
subset of most sensitive nodes involved in a pathway between these two nodes. 
Furthermore we apply to this subset of nodes the REduced GOogle MAtriX 
(REGOMAX) algorithm developed in \cite{greduced,politwiki} and obtain 
in this way an effective Google matrix description between nodes
of the found pathway. 

In general the REGOMAX algorithm determines effective interactions 
between selected nodes of a certain relatively small subset 
embedded in a global huge network. Its efficiency was 
recently demonstrated for the Wikipedia networks
of politicians \cite{politwiki} and world universities \cite{wrwu2017},
SIGNOR network of protein-protein interactions \cite{zinovyev} and
multiproduct world trade network of UN COMTRADE \cite{wtn2019}.

In this work our aim is to provide a first illustration of the efficiency
of the LIRGOMAX algorithm combined with the reduced 
Google matrix analysis. Due to this we restrict in this work 
our considerations to the analytical description 
of the LIRGOMAX algorithm and the illustration of its application 
to two cases from the English Wikipedia network of 2017. 

The paper is constructed as follows:
in Section 2 we provide the analytical description
of the LIRGOMAX algorithm complemented by a brief description of
the Google matrix construction and the REGOMAX algorithm,
in Section 3 we present certain results for two examples of the 
Wikipedia network, the discussion of results is given in Section 4.
Additional data are also available at \cite{ourwebpage}.

\section{Theory of a weakly perturbed 
Google matrix}

\subsection{Google matrix construction}

We first briefly 
remind the general construction of the Google matrix $G$ from 
a direct network of $N$ nodes. For this one first computes 
the adjacency matrix $A_{ij}$ with elements $1$ 
if node $j$ points to node $i$ and zero otherwise. 
The matrix elements of $G$ have the usual form 
$G_{ij} = \alpha S_{ij} + (1-\alpha) / N$ \cite{brin,meyer,rmp2015},
where $S$ is the matrix of Markov transitions with elements  
$S_{ij}=A_{ij}/k_{out}(j)$ and $k_{out}(j)=\sum_{i=1}^{N}A_{ij}\neq 0$ 
being the  out-degree of node $j$ 
(number of outgoing links) or  $S_{ij}=1/N$ if $j$ 
has no outgoing links (dangling node). 
The parameter $0< \alpha <1$ is the damping factor
with the usual value $\alpha=0.85$ \cite{meyer} used here.
We note that for the range $0.5 \leq \alpha \leq 0.95$
the results are not sensitive to $\alpha$ \cite{meyer,rmp2015}. 
This corresponds to a model of a random surfer who follows with probability 
$\alpha$ at random one of the links available from the actual node 
or jumps with probability $(1-\alpha)$ to an arbitrary other node in 
the network. 

The right PageRank eigenvector
of $G$ is the solution of the equation $G P = \lambda P$
for the unit eigenvalue $\lambda=1$ \cite{brin,meyer}. The PageRank 
$P(j)$ values represent positive probabilities to find a random surfer 
on a node $j$ ($\sum_j P(j)=1$). 
All nodes can be ordered by decreasing probability $P$ 
numbered by the PageRank index $K=1,2,...N$ with a maximal probability at 
$K=1$ and minimal at $K=N$. The numerical computation 
of $P(j)$ is done efficiently with the PageRank iteration
algorithm described in \cite{brin,meyer}.

It is also useful to consider
the original network with inverted direction of links.
After inversion the Google matrix $G^*$ is constructed via
the same procedure (using the transposed adjacency matrix) and its 
leading eigenvector $P^*$, determined by $G^* P^*= P^*$, 
is called CheiRank \cite{cheirank} (see also \cite{rmp2015}).
Its values $P^*(j)$ can be again ordered
in decreasing order resulting in the CheiRank index $K^*$
with highest value of $P^*$ at $K^*=1$ and smallest values at $K^*=N$.
On average, the high values of $P$ ($P^*$) correspond to nodes
with many ingoing (outgoing) links \cite{meyer,rmp2015}.

\subsection{Reduced Google matrix algorithm}

The REGOMAX method is described in detail 
in $\;\;\;\;$ \cite{greduced,politwiki,zinovyev,wrwu2017}. For a given 
relatively small subset of $N_r\ll N$ nodes it allows to 
compute efficiently a ``reduced Google matrix'' 
$\GR$ of size $N_r \times N_r$ that captures
the full contributions of direct and indirect pathways appearing 
in the full Google matrix $G$ between the $N_r$ selected nodes of interest. 
The PageRank vector $P_r$ of $\GR$ coincides with the full PageRank vector 
projected on the subset of nodes, 
up to a constant multiplicative factor due to the sum normalization. 
The mathematical computation of $\GR$ provides 
a decomposition of $\GR$ into matrix components that clearly distinguish 
direct from indirect interactions: 
$\GR = \Grr + \Gpr + \Gqr$ \cite{politwiki}.
Here $\Grr$ is given by the direct links between the selected 
$N_r$ nodes in the global $G$ matrix with $N$ nodes. 
$\Gpr$ is a rank one matrix whose columns are rather close (up to constant 
factor) to the reduced PageRank vector $P_r$. 
Even though the numerical weight of $\Gpr$ is typically quite large 
it does not give much new interesting information about the reduced 
effective network structure. 

The most interesting role is played by $\Gqr$, which takes 
into account all indirect links between
selected nodes happening due to multiple pathways via 
the global network nodes $N$ (see~\cite{greduced,politwiki}).
The matrix  $\Gqr=\Gqrd + {\Gqr}^{(nd)}$ has diagonal ($\Gqrd$)
and non-diagonal (${\Gqr}^{(nd)}$) parts with ${\Gqr}^{(nd)}$
describing indirect interactions between selected nodes.
The exact formulas and the numerical algorithm for an efficient numerical 
computation of all three components of $\GR$ are given 
in \cite{greduced,politwiki}. It is also useful 
to compute the weights $\WR$, $\Wpr$,
$\Wrr$, $\Wqr$
of $\GR$ and its 3 matrix components $\Gpr$, $\Grr$, $\Gqr$
given by the sum of all its elements divided by the matrix size $N_r$. 
Due to the column sum normalization of $\GR$ we obviously have 
$\WR=\Wrr+\Wpr+\Wqr=1$. 

\subsection{General model of linear response}

We consider a Google matrix $G(\eps)$ (with non-negative matrix elements 
satisfying the usual column sum normalization) depending on a small parameter 
$\eps$ and a general stochastic process $P(t+1)=G(\eps)\,F(\eps,P(t))$
where $F(\eps,P)$ is a general function on $\eps$ and $P$ which does NOT need 
to be linear in $P$. (Here $P(t)$ denotes a time dependence of the vector $P$; 
below and for the rest of this paper we will use the notation 
$P(j)$ for the $j$th component of the vector $P$).

Let $E^T=(1,\ldots,1)$ be the usual vector with unit 
entries. Then the condition of column sum normalization of $G(\eps)$ reads 
$E^T G(\eps)=E^T$. The function $F(\eps,P)$ should satisfy the 
condition $E^T F(\eps,P)=1$ if $E^T P=1$. At $\eps=0$ we also require that 
$F(0,P)=P$, i.e. $F(0,P)$ is the identity operation on $P$. 
We denote by $G_0=G(0)$ the Google matrix at $\eps=0$ and by $P_0$ its 
PageRank vector such that $G_0\,P_0=P_0$ with $E^T P_0=1$. 
We denote by $P$ the more general, $\eps$-dependent, solution of 
\begin{equation}
\label{eq_Psol1}
P=G(\eps)\,F(\eps,P)\quad,\quad E^T\,P=1\ .
\end{equation}

\subsubsection{Pump model}

As a first example we present the {\em Pump model} to 
model an injection- and absorption scheme. For this 
we choose for the Google matrix simply $G(\eps)=G_0$ 
(i.e. no $\eps$-dependence for $G$) and
\begin{equation}
\label{eq_Finject}
F(\eps,P)=\frac{({\bf 1}+\eps D)P}{E^T[({\bf 1}+\eps D)P]}=
\frac{({\bf 1}+\eps D)P}{1+\eps\,e(P)}
\end{equation}
with $e(P)=E^T D P$ and $D$ being a diagonal matrix with entries $D_j$ 
which are mostly zero and with few positive values $D_j>0$ for 
nodes $j$ with injection and few negative values $D_j<0$ for nodes $j$ with 
absorption. 
The non-vanishing diagonal entries of $D_j$ should be comparable (a global 
scaling factor can be absorbed in a redefinition of the parameter $\eps$) 
and we have $e(P)=\sum_{D_j\neq 0} D_j P(j)$. 
Physically, we multiply each entry $P(j)$ by the factor $1+\eps D_j$ 
(which is unity for most nodes $j$) and then we sum-normalize this vector to 
unity before we apply the Google matrix $G_0$ to it. 

\subsubsection{PageRank sensitivity}

As a second example we consider the PageRank sensitivity. For this we 
fix a pair $(i,j)$ of indices and multiply the matrix element 
$(G_0)_{ij}$ by $(1+\eps)$ and then we sum-normalize the column $j$ to unity 
which provides the $\eps$-dependent Google matrix $G(\eps)$. For the 
function $F(\eps,P)$ we simply choose the identity operation: $F(\eps,P)=P$. 
In a more explicit formula we have:
\begin{equation}
\label{eq_Gsensdef}
\forall_{k,l}\qquad
G_{kl}(\eps)=\frac{(1+\eps\,\delta_{ki}\delta_{lj})\,(G_0)_{kl}}
{1+\eps\,\delta_{lj}\,(G_0)_{ij}}
\end{equation}
where $\delta_{ki}=1$ (or $0$) if $k=i$ (or $k\neq i$). 
Note that the denominator is either $1$ if $l\neq j$ 
or the modified column sum $1+\eps\,(G_0)_{ij}$ of column $j$ if $l=j$. 
Then the sensitivity $D_{(j\to i)}(k)$ is defined as:
\begin{equation}
\label{eq_sensdef}
D_{(j\to i)}(k)=\frac{P(k)-P_0(k)}{\eps P_0(k)}
\end{equation}
where $P$ is the $\eps$-dependent PageRank of $G(\eps)$ computed in 
the usual way. We expect that this quantity has a well defined limit 
if $\eps\to 0$ but equation (\ref{eq_sensdef}) is numerically not 
very precise for very small values of $\eps$ due to the effect of loss 
of precision. Below we present a method to compute the sensitivity in a 
precise way in the limit $\eps\to 0$.

Examples of the sensitivity analysis, using directly (\ref{eq_sensdef}), 
were considered for the reduced Google matrix of sets of
Wikipedia and other networks (see e.g. \cite{wrwu2017,wtn2019}).

\subsection{Linear response}

\subsubsection{General scheme}

One can directly numerically determine the $\eps$-dependent 
solution $P(\eps)$ of (\ref{eq_Psol1}) 
for some small but finite value of $\eps$ (by iterating 
$P^{(n+1)}=G(\eps),F(\eps,P^{(n)})$ with some suitable initial vector 
$P^{(0)}$) and compute the quantity 
\begin{equation}
\label{eq_Pdelta1}
\Delta P(\eps)=\frac{P(\eps)-P(0)}{\eps}\ .
\end{equation}
We expect that $\Delta P(\eps)$ has a finite well defined limit if 
$\eps\to 0$. 
However, its direct numerical computation by (\ref{eq_Pdelta1}) is subject
to numerical loss of precision if $\eps$ is too small. 
In the following, we will 
present a different scheme to compute $\Delta P$ which is numerically more 
accurate and stable and that we call linear response of Google matrix. 
For this we expand $G(\eps)$ and $F(\eps,P)$ up to order $\eps^1$ 
(neglecting terms $\sim \eps^2$ or higher):
\begin{equation}
\label{eq_expand1}
G(\eps)=G_0+\eps G_1+\ldots \quad,\quad F(\eps,P)=P+\eps F_1(P)+\ldots
\end{equation}
Furthermore we write 
\begin{equation}
\label{eq_Pexpand1}
P(\eps)=P_0+\eps\,P_1+\ldots\ .
\end{equation}
The usual sum-normalization conditions for the first order corrections 
read as~: 
\begin{equation}
\label{eq_sum1}
E^T G_1=0\quad,\quad E^T F_1(P)=0\quad,\quad
E^T P_1=0
\end{equation}
if $E^T P = E^T P_0= 1$. These conditions imply that $P_1$ and also 
$F_1(P)$ belong to the subspace ``bi-orthogonal'' to the PageRank, i.e. 
orthogonal to the left leading eigenvector of $G_0$ which is just 
the vector $E^T$. 

Inserting (\ref{eq_expand1}) and 
(\ref{eq_Pexpand1}) into (\ref{eq_Psol1}) we obtain (up to order 
$\eps^1$):
\begin{equation}
\label{eq_Ppert1}
P=P_0+\eps P_1=
G_0\,P_0+\eps\Bigl[G_0\,P_1+G_1\,P_0+G_0\,F_1(P_0)\Bigr]\ .
\end{equation}
Comparing the terms of order $\eps^0$ one obtains the usual unperturbed 
PageRank equation $P_0=G_0\,P_0$. The terms of order $\eps^1$ provide 
an inhomogeneous PageRank equation of the type~:
\begin{equation}
\label{eq_inhom1}
P_1=G_0\,P_1+V_0\quad,\quad V_0=G_1\,P_0+G_0\,F_1(P_0)\ .
\end{equation}
The solution $P_1$ of this equation is just the limit of (\ref{eq_Pdelta1}):
\begin{equation}
\label{eq_P1lim}
P_1=\lim_{\eps\to 0} \Delta P(\eps)=
\lim_{\eps\to 0} \frac{P(\eps)-P(0)}{\eps}\ .
\end{equation}
To solve numerically (\ref{eq_inhom1}) we first determine the 
unperturbed PageRank $P_0$ of $G_0$ in the usual way and compute $V_0$ which 
depends on $P_0$. Then we iterate the equation:
\begin{equation}
\label{eq_P1iterate}
P_1^{(n+1)}=G_0\,P_1^{(n)}+V_0
\end{equation}
where for the initial vector we simply choose $P_1^{(0)}=0$. This iteration 
converges with the same speed as the usual PageRank algorithm 
versus the vector $P_1$ and it is numerically more accurate than the 
finite difference $\Delta P(\eps)$ at some finite value of $\eps$. 

We remind that $E^T V_0=\sum_j V_0(j)=0$ and also $E^T G_0=E^T$. 
Therefore if at a given iteration step the vector $P_1^{(n)}$ satisfies 
the condition $E^T P_1^{(n)}=0$ we also have 
$E^T P_1^{(n+1)}=E^T G_0 P_1^{(n)}+E^T V_0= E^T P_1^{(n)}=0$. 

Therefore 
the conditions (\ref{eq_sum1}) are satisfied by the iteration equation 
(\ref{eq_P1iterate}) at least on a theoretical/mathematical level. However, 
rounding errors may produce slight errors in the conditions (\ref{eq_sum1}) 
and since such numerical errors contain a contribution in the direction of 
the unperturbed 
PageRank vector $P_0$, corresponding to the eigenvector of $G_0$ 
with maximal eigenvalue, they do not disappear during the iteration and might 
even (slightly) increase with $n$. Therefore, due to purely numerical 
reasons, it is useful to remove such contributions by a projection
after each iteration step of 
the vector $P_1^{(n+1)}$ on the subspace bi-orthogonal to the PageRank by:
\begin{equation}
\label{eq_projectQ}
P_1^{(n+1)}\to Q\left(P_1^{(n+1)}\right)\quad,\quad Q(X)=X-(E^T X)\,P_0
\end{equation}
where $Q(X)$ is the projection operator applied on a vector $X$. 
It turns out that such an additional projection step indeed 
increases the quality and accuracy of the convergence of 
(\ref{eq_P1iterate}) but even without it the iteration (\ref{eq_P1iterate}) 
converges numerically, however with a less accurate result. 

It is interesting to note that one can ``formally'' solve (\ref{eq_P1iterate})
by:
\begin{equation}
\label{eq_P1solve}
P_1=\sum_{n=0}^\infty G_0^n\,V_0=\frac{\bf 1}{{\bf 1}-G_0}\,V_0
\end{equation}
which can also be found directly from the first equation in (\ref{eq_inhom1}). 
Strictly speaking the matrix inverse $({\bf 1}-G_0)^{-1}$ does not exist 
since $G_0$ has always one eigenvalue $\lambda=1$. However, since 
$E^T\,V_0=0$, the vector $V_0$, when expanded in the basis of (generalized) 
eigenvectors of $G_0$, does not contain a contribution of $P_0$ which is 
the eigenvector for $\lambda=1$ such that the expression (\ref{eq_P1solve}) 
is actually well defined. From a numerical point of view a different 
scheme to compute $P_1$ would be to 
solve directly the linear system of equations $({\bf 1}-G_0)\,P_1=V_0$ where 
the first (or any other suitable) equation of this system is replaced by 
the condition $E^T P_1=0$ resulting in a linear system with a well defined 
unique solution. Of course, such a direct computation is limited to modest 
matrix dimensions $N$ such that a full matrix inversion is possible 
(typically $N$ being a few 
multiples of $10^4$) while the iterative scheme (\ref{eq_P1iterate}) 
is possible for rather large matrix dimensions such that the usual PageRank 
computation by the power method is possible. For example for the 
English Wikipedia edition of 2017 with $N\approx 5\times 10^6$ the 
iterative computation of $P_1$ using (\ref{eq_P1iterate}) 
takes typically $2-5$ minutes on a recent single processor core 
(e.g.: Intel i5-3570K CPU) without any use of parallelization 
once the PageRank $P_0$ is known. (The computation of $P_0$ by the usual 
power method takes roughly the same time.) 

\subsubsection{Application to the pump model} 

For the injection- and absorption scheme 
we can compute $F_1(P)$ from (\ref{eq_Finject}) as:
\begin{eqnarray}
\label{eq_F1P_compute}
F(\eps,P)&=&({\bf 1}-\eps\,e(P)+\ldots)(P+\eps D P)
\\
\nonumber
&=& P+\eps\,[P-(E^T D P)P]+\ldots\ .
\end{eqnarray}
Here the term $\sim \eps^1$ is just the projection of $DP$ to the 
subspace bi-orthogonal to $P$. This projection is the manifestation 
in first order in $\eps$ of the renormalization used in 
(\ref{eq_Finject}). 

Furthermore, since for the injection- and absorption scheme we 
also have $G_1=0$, the vector $V_0$ in (\ref{eq_inhom1}) 
and (\ref{eq_P1iterate}) becomes:
\begin{eqnarray}
\nonumber
V_0&=&G_0\,F_1(P_0)=G_0\,Q(D\,P_0)\\
\label{eq_V0pump}
&=&G_0 D\,P_0-(E^T D P_0)\,G_0\,P_0\\
\nonumber
&=&G_0 D\,P_0-(E^T G_0\,D P_0)\,P_0=Q(G_0\,D\,P_0)
\end{eqnarray}
with $Q$ being the projector 
given in (\ref{eq_projectQ}). Here we have used that $E^T\,G_0=E^T$ 
and $G_0\,P_0=P_0$. This small calculation also shows that the projection 
operator can be applied before or after multiplying $G_0$ to $D\,P_0$.

\subsubsection{Application to the sensitivity}

In this case we have $F(\eps,P)=P$ such that $F_1(P)=0$ and we have to 
determine $G_1$ from the expansion $G(\eps)=G_0+\eps G_1+\ldots$. 
Expanding (\ref{eq_Gsensdef}) up to first order in $\eps$ we obtain:
\begin{equation}
\label{eq_G1sens}
\forall_{kl}\qquad (G_1)_{kl}=\delta_{ki}\delta_{lj}\,(G_0)_{kl}-
\delta_{lj}\,(G_0)_{ij}
\end{equation}
where $(i,j)$ is the pair of indices for which we want to compute the 
sensitivity. Using $G_1$ we compute $V_0=G_1\,P_0$ and solve 
the inhomogeneous PageRank equation (\ref{eq_inhom1}) iteratively as described 
above to obtain $P_1$. Once $P_1$ is know we can compute the sensitivity 
from~:
\begin{equation}
\label{eq_sensexact}
D_{(j\to i)}(k)=\frac{P_1(k)}{P_0(k)}\ .
\end{equation}
We note that equation (\ref{eq_sensexact}) is numerically accurate 
and corresponds to the exact limit $\eps\to 0$ while 
(\ref{eq_sensdef}) is numerically not very precise and requires 
a finite small value of $\eps$. 

\begin{table}
\begin{center}
\caption{Top 20 nodes of strongest negative values of 
$P_1$ (index number $i=1,\ldots,20$) and top 20 nodes of strongest 
positive values of $P_1$ (index number $i=21,\ldots,40$) 
with $P_1$ being created as the linear response of PageRank of English 
Wikipedia 2017 PageRank with injection (or pumping)  at {\it University of Cambridge} 
and absorption at {\it Harvard University}; $K_L$ is the ranking index 
obtained by ordering $|P_1|$ and $K$ is the usual PageRank index
obtained by  ordering the PageRank probability $P_0$ of the global 
network with $N$ nodes. 
}
{\relsize{-2}
\label{table1}
\begin{tabular}{rrrl}
\hline
$i$ & $K_L$ & $K$ &Node name \Tstrut\Bstrut\\
\hline\Tstrut
1 & 1 & 129 & Harvard University \\
2 & 2 & 1608 & Cambridge, Massachusetts \\
3 & 4 & 1 & United States \\
4 & 5 & 296 & Yale University \\
5 & 6 & 4617 & Harvard College \\
6 & 7 & 62115 & Harvard Yard \\
7 & 10 & 104359 & Harvard Museum of Natural History \\
8 & 11 & 415 & National Collegiate Athletic Ass. \\
9 & 12 & 52 & The New York Times \\
10 & 13 & 75 & American Civil War \\
11 & 14 & 7433 & Harvard Medical School \\
12 & 15 & 73 & American football \\
13 & 16 & 20994 & Charles River \\
14 & 17 & 50 & Washington, D.C. \\
15 & 18 & 23901 & Harvard Divinity School \\
16 & 19 & 436 & Massachusetts Institute of Tech. \\
17 & 20 & 88022 & President and Fellows of Harvard Col. \\
18 & 21 & 128 & Boston \\
19 & 22 & 42608 & The Harvard Crimson \\
\Bstrut 20 & 23 & 42259 & Harvard Square \\
\hline\Tstrut
21 & 3 & 229 & University of Cambridge \\
22 & 8 & 15 & England \\
23 & 9 & 1414 & Cambridge \\
24 & 69 & 1842 & Trinity College, Cambridge \\
25 & 253 & 6591 & St John's College, Cambridge \\
26 & 254 & 7022 & King's College, Cambridge \\
27 & 256 & 285 & Order of the British Empire \\
28 & 257 & 6 & United Kingdom \\
29 & 258 & 33256 & Newnham College, Cambridge \\
30 & 260 & 316 & Church of England \\
31 & 262 & 238 & The Guardian \\
32 & 263 & 21569 & Clare College, Cambridge \\
33 & 264 & 4656 & Durham University \\
34 & 265 & 191614 & Regent House \\
35 & 266 & 3814 & Chancellor (education) \\
36 & 267 & 16193 & Gonville and Caius Col. Cambridge \\
37 & 270 & 25165 & E. M. Forster \\
38 & 274 & 1650 & Archbishop of Canterbury \\
39 & 277 & 2076 & Fellow \\
\Bstrut 40 & 278 & 73538 & Colleges of the Univ. of Cambridge \\
\hline
\end{tabular}
}
\end{center}
\end{table}


\subsection{LIRGOMAX combined with REGOMAX}

We consider the pump model described above and we take two particular 
nodes $i$  with injection  and $j$ with absorption. For the diagonal matrix $D$ we choose
$D_i=1/P_0(i)$ and $D_j=-1/P_0(j)$ where $P_0$ is the PageRank of the 
unperturbed network and all other values $D_k=0$. 
In this way we have $e(P_0)=E^T D\,P_0=D_i\,P_0(i)+D_j\,P_0(j)=0$.
Due to this
the renormalization denominator in (\ref{eq_Finject}) is simply unity 
and all excess probability provided by the injection at node $i$ will be 
exactly absorbed by the absorption at node $j$. We insist that this is only 
due to our particular choice for the matrix $D$ 
and concerning the numerical procedure one can also 
choose different values of $D_i$ or $D_j$ with $e(P_0)\neq 0$ 
(which would result in some global excess probability which 
would be artificially injected or absorbed due the normalization denominator 
in (\ref{eq_Finject}) being different from unity). 

Using the above values of $D_i$ and $D_j$ we compute the vector 
$V_0=G_0\,D\,P_0=G_0\,W_0$ (since $E^T D\,P_0=0$) where $W_0=D\,P_0$ 
is a vector with only two non-zero components $W_0(i)=1$ and $W_0(j)=-1$. 
Therefore for all $k$ we have $V_0(k)=(G_0)_{ki}-(G_0)_{kj}$. According to the 
above theory we know that $V_0$ and $W_0$ are orthogonal to $E^T$, i.e. 
$E^T V_0=E^T W_0=0$ or more explicitely $\sum_k V_0(k)=\sum_k W_0(k)=0$. 
For $W_0$ the last equality is obvious and the first one is due to the column 
sum normalization of $G_0$ meaning that 
$\sum_k (G_0)_{ki}=\sum_k (G_0)_{kj}=1$.
Using the expression $V_0(k)=(G_0)_{ki}-(G_0)_{kj}$ 
we determine the solution of the linear response 
correction to the PageRank $P_1$ by solving iteratively the 
inhomogeneous PageRank equation (\ref{eq_inhom1}) as described above. 
The vector $P_1$ has real positive and negative entries also 
satisfying the condition $\sum_k P_1(k)=0$. 
Then we determine the 20 top nodes with strongest negative values of $P_1$ and 
further 20 top nodes with strongest positive values of $P_1$ which 
constitute a subset of 40 nodes which are the most significant
nodes participating in the pathway between the pumping node $i$ and 
absorbing node $j$. 

Using this subset we then apply the REGOMAX algorithm 
to compute the reduced Google matrix and its components which 
are analyzed in a similar way as in \cite{politwiki}. 
The advantage of the application of LIRGOMAX at the initial stage 
is that it provides an automatic procedure to determine an interesting 
subset of nodes related to the pumping between nodes $i$ and $j$ instead 
of using an arbitrary heuristic choice for such a subset. 

The question arises if the initial two nodes $i$ and $j$ belong themselves 
to the subset of nodes with largest $P_1$ entries (in modulus). From a 
physical point of view we indeed expect that this is generically the 
case but there is no simple mathematical argument for this. In particular 
for nodes with a low PageRank ranking and zero or few incoming links this is 
probably not the case. 
However, concerning the two examples which we will 
present in the next section both 
initial nodes $i$ and $j$ are indeed present in the selected subset and 
even with rather top positions in the ranking (provided by ordering $|P_1|$).

\section{LIRGOMAX for Wikipedia network}

As a concrete example we illustrate the application of LIRGOMAX algorithm
to the English Wikipedia network of 2017 
(network data available at \cite{wiki2017}).
This network contains $N=5416537$ nodes,
corresponding to article titles, and $N_i = 122232932$
directed hyperlinks between nodes. 
Previous applications of the REGOMAX algorithm for
the Wikipedia networks of years 2013 and 2017 are
described in \cite{politwiki,wrwu2017}.

\subsection{Case of  pathway Cambridge - Harvard Universities}

As a first example of the application of the combined LIRGOMAX and REGOMAX 
algorithms we select two articles (nodes) of the Wikipedia network with 
pumping at {\it University of Cambridge} and absorption at {\it Harvard University}.
The global PageRank indices of these two nodes are
$K=229$ (PageRank probability $P_0(229)= 0.0001078$)
and $K=129$ (PageRank probability $P_0(129)=  0.0001524$). 
As described above we chose the diagonal matrix $D$ as 
$D(229)=1/P_0(229)$ and $D(129)=-1/P_0(129)$ (other diagonal entries of $D$ 
are chosen as zero) 
and determine the vector $V_0$ used for the computation of $P_1$
(see (\ref{eq_P1iterate})) by $V_0=G_0\,W_0$ 
where the vector $W_0=D\,P_0$ has the nonzero components $W_0(229) = 1$ and 
$W_0(129) = -  1$. Both $W_0$ and $V_0$ are orthogonal to the left 
leading eigenvector $E^T=(1,\ldots,1)$ of $G_0$ according to the theory 
described in the last section. 

\begin{figure}
\begin{center}
\includegraphics[width=0.48\textwidth]{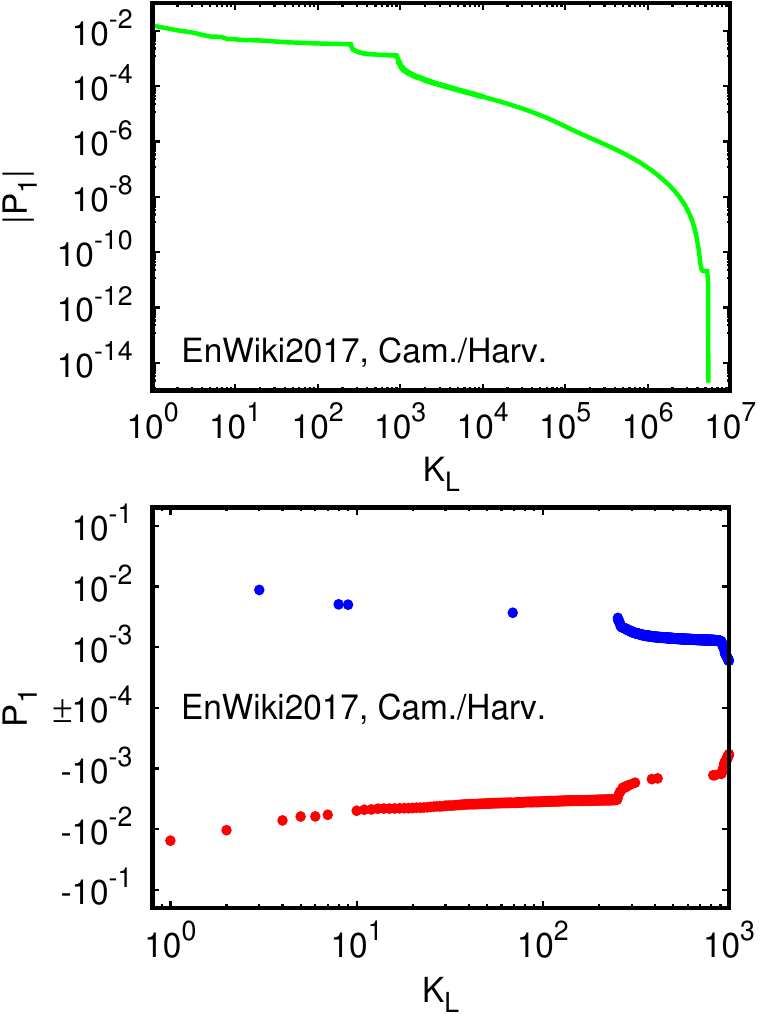}
\caption{Linear response vector $P_1$ of PageRank for the English 
Wikipedia 2017  with injection (or pumping) at {\it University of Cambridge} 
and absorption at {\it Harvard University}. Here $K_L$ is the ranking index 
obtained by ordering $|P_1|$ from maximal value at $K_L=1$
down to its minimal value. Top panel shows $|P_1|$ versus $K_L$ in 
a double logarithmic representation for all $N$ nodes.
Bottom panel shows a zoom of $P_1$ versus $K_L$ for $K_L \le 10^3$ 
in a double logarithmic representation with sign; 
blue data points correspond 
to $P_1>0$ and red data points to $P_1<0$. 
}
\label{fig1}
\end{center}
\end{figure}

The subset of 40 most affected nodes with 20 strongest negative and 20 
strongest positive values of the linear response correction $P_1$ 
to the initial PageRank $P_0$ are given in Table~\ref{table1}.
We order these 40 nodes by the index $i=1, \ldots, 20$ for the first 20 most 
negative $P_1$ values and then $i=21, \ldots, 40$ for the 
most positive $P_1$ values. The index $K_L$ is obtained by ordering 
$|P_1|$ for all $N\approx 5\times 10^6$ network nodes. 
The table also gives the PageRank index 
$K$ obtained by ordering $P_0$. 
The first 4 positions in $K_L$ are taken by
{\it Harvard University; Cambridge, Massachusetts; 
University of Cambridge; United States}.
Thus, even if the injection is made for
{\it University of Cambridge} the strongest response appears for
{\it  Harvard University; Cambridge, Massachusetts}
and only then for {\it University of Cambridge} ($K_L=1,2,3$).
We attribute this to nontrivial flows existing in the global directed
network. This shows that the linear response approach
provides rather interesting information
about the sensitivity and interactions of nodes on directed networks.
We will see below for other examples that the top nodes
of the linear response vector $P_1$ can have rather unexpected features.

In general the most sensitive nodes of Table~\ref{table1}
are rather natural. They represent countries, cities
and other administrative structures related to the two universities.
Other type of nodes are 
{\it Yale University, The New York Times, American Civil War}
for Harvard U and
{\it Church of England, The Guardian, Durham University}
for U Cambridge (in addition to many Colleges presented in the list) 
corresponding to closest other universities and also newspapers appearing 
on the pathway between the pair of selected nodes.

Of course, the linear response vector $P_1$
extends on all $N$ nodes of the global network.
We show its dependence on the ordering index 
$K_L$ in Figure~\ref{fig1}. Here the top
panel represents the decay of $|P_1|$ with $K_L$
and the bottom panel shows the decay of negative and 
and positive $P_1$ values for $K_L \leq 10^3$.
We note that among top 100 values of $K_L$
there are only 4 nodes related to U Cambridge
with positive $P_1$ values while all
other values of $P_1$ are negative being 
related to Harvard U.
This demonstrates a rather different structural influence between 
these two universities.

\begin{figure}
\begin{center}
\includegraphics[width=0.48\textwidth]{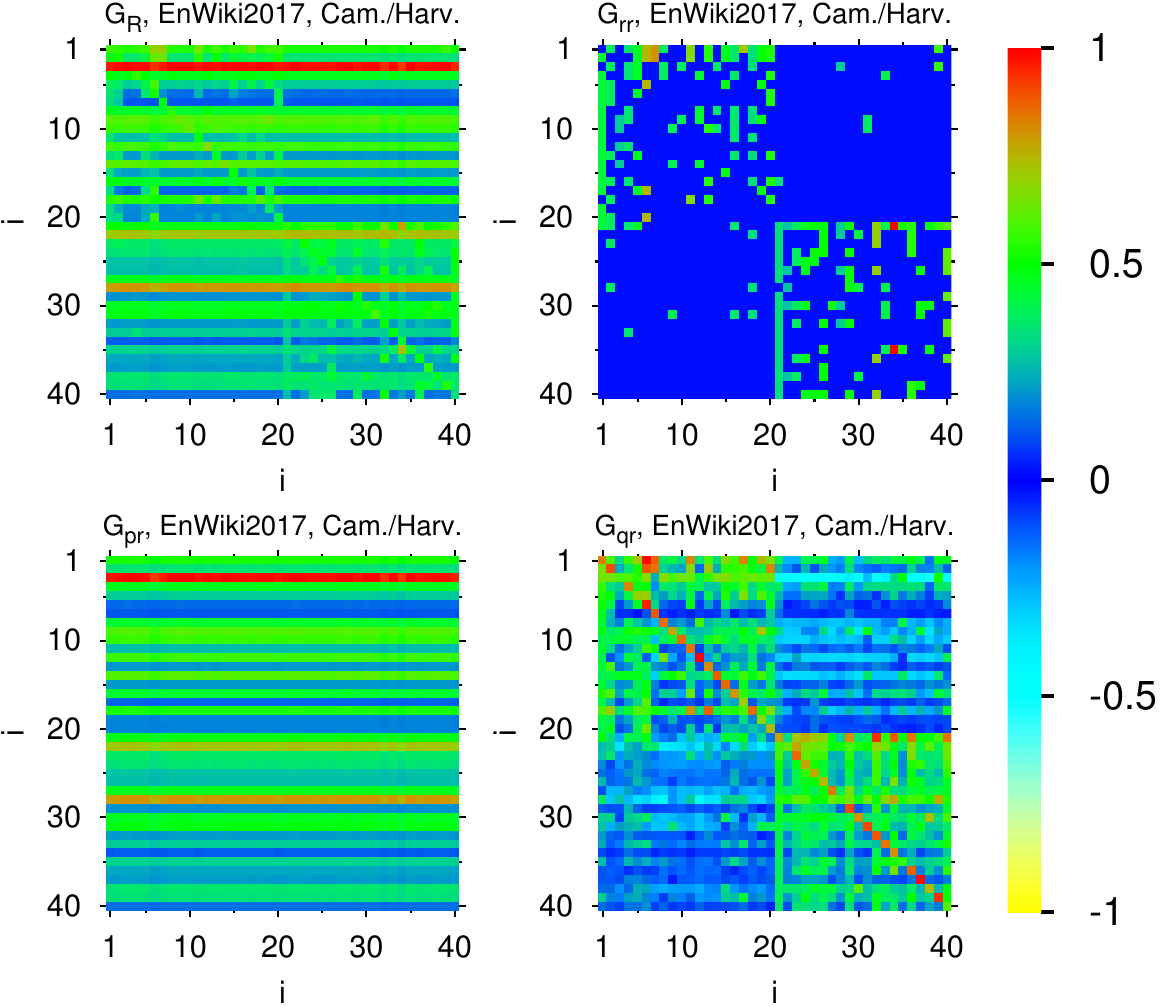}
\caption{Reduced Google matrix components $\GR$, $\Gpr$, 
$\Grr$ and $\Gqr$ for the English Wikipedia 2017 network and the 
subgroup of nodes given in Table~\ref{table1} corresponding to 
injection at {\it University of Cambridge} 
and absorption at {\it Harvard University} (see text for explanations). The 
axis labels correspond to the index number $i$ used in Table~\ref{table1}.
The relative weights of these components are $\Wpr=0.920$, $\Wrr=0.036$, 
 and $\Wqr=0.044$. Note that elements of $G_{qr}$ may 
be negative. The values of the color bar correspond to 
$\sgn(g)(|g|/\max|g|)^{1/4}$ where $g$ is the shown 
matrix element value. The exponent $1/4$ amplifies 
small values of $g$ for a better visibility. 
}
\label{fig2}
\end{center}
\end{figure}

After the selection of 40 most significant nodes of the pathway
between both universities (see Table~\ref{table1})
we apply the REGOMAX algorithm
which determines all matrix elements of Markov transitions
between these 40 nodes including all direct and indirect pathways
via the huge global Wikipedia network with 5 million nodes.

The reduced Google matrix $\GR$ and its three components
$\Gpr$, $\Grr$, $\Gqr$ are shown in Figure~\ref{fig2}.
As discussed above the weight $\Wpr=0.920$ of $\Gpr$
is close to unity and its matrix structure
is rather similar to the one of $\GR$ 
with  strong transition lines of matrix elements
corresponding to {\it United States} at top PageRank index $K=1\ (i=3,\ K_L=4)$
and {\it United Kingdom} at $K=6\ (i=28,\ K_L=257)$.
The weights $\Wrr =0.036 $, $\Wqr = 0.044$ of $\Grr$, $\Gqr$ are significantly
smaller. These values are similar to those obtained in
the REGOMAX analysis of politicians and universities in Wikipedia networks
\cite{politwiki,wrwu2017}. Even if the weights of these matrix components
are not large they represent the most interesting and
nontrivial direct ($\Grr$) and indirect ($\Gqr$)  interactions
between the selected 40 nodes. The image of $\Grr$ in Figure~\ref{fig2}
shows that the direct links between the U Cambridge block of nodes
(with index $21 \leq i \leq 40$ in Table~\ref{table1})
and the Harvard U block of nodes (with index $1 \leq i \leq 20$ in 
Table~\ref{table1}) are rather rare and relatively weak
while the links within each block 
are multiple and relatively strong. This confirms the appropriate selection 
of nodes in each block provided by the LIRGOMAX algorithm.

\begin{figure}
\begin{center}
\includegraphics[width=0.48\textwidth]{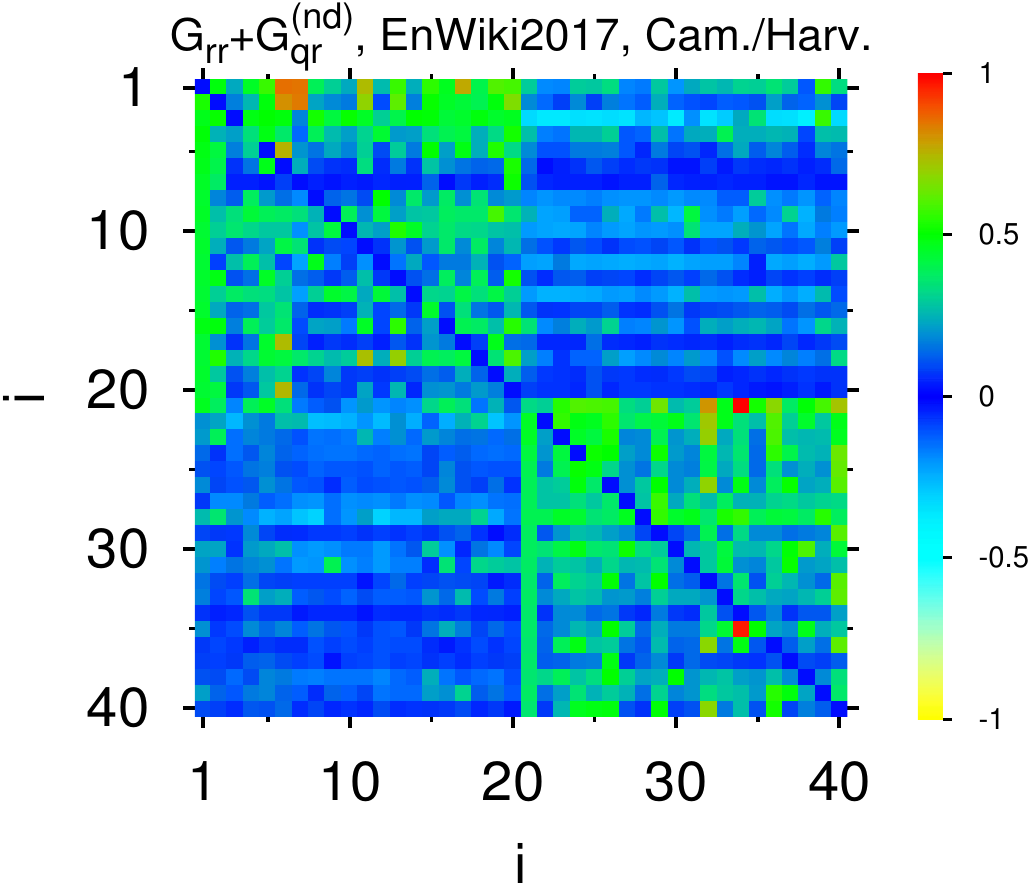}
\caption{Same as in Fig.~\ref{fig2} but for the matrix $G_{rr}+G_{qr}^{(nd)}$, 
where $G_{qr}^{(nd)}$ is obtained from $G_{qr}$ by putting its diagonal 
elements at zero; the weight of these two components is 
$W_{rr+qrnd}=0.066$.
}
\label{fig3}
\end{center}
\end{figure}

The matrix elements of the sum of two components
$\Grr+{\Gqr}^{(nd)}$ (component $\Gqr$ is taken without diagonal elements)
are shown in Figure~\ref{fig3}. We note that some elements
are negative which is not forbidden since only the sum of all three
components given by $\GR$ should have positive matrix elements.
However, the negative values are rare and relatively small
compared to the values of positive matrix elements.
Thus the minimal value is $-0.00216$ for the transition from {\it Church of England}
to {\it United States} while other typical 
negative values are smaller by a factor 5-10. For comparison, the maximal
value of positive element is
$ 0.1135$ from {\it Regent House} to {\it  University of Cambridge}
and there are many other positive elements of the order of $0.03$.
Thus we consider that the negative elements play no significant role.
A similar conclusion was also obtained for the interactions of politicians 
and universities in \cite{politwiki,wrwu2017}.

\begin{figure}
\begin{center}
\includegraphics[width=0.48\textwidth]{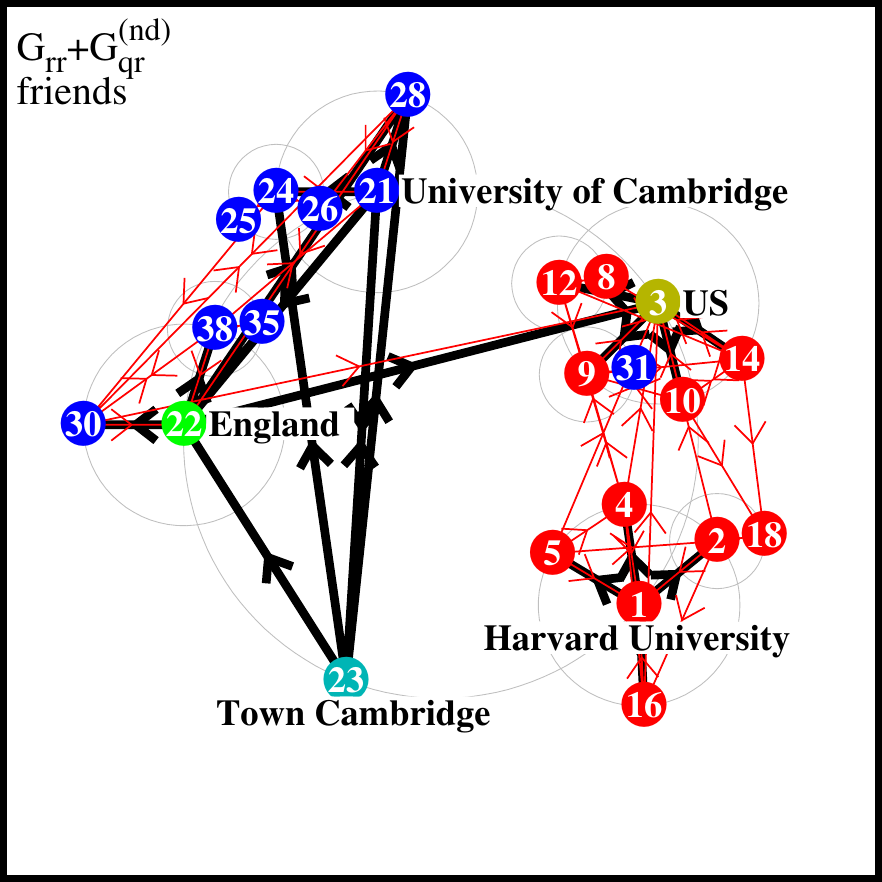}
\caption{Network of friends for the subgroup of nodes given in 
Table~\ref{table1} 
corresponding to injection at {\it University of Cambridge} 
and absorption at {\it Harvard University} constructed from the matrix 
$G_{rr}+G_{qr}^{(nd)}$ using 4 top (friends) links 
per column (see text for explanations). The numbers used as labels for 
the different nodes correspond to the index $i$ of Table~\ref{table1}. 
}
\label{fig4}
\end{center}
\end{figure}

From Figure~\ref{fig3} we see that for $G_{rr}+G_{qr}^{(nd)}$
the strongest interactions are also inside each university block.
However, there are still some significant links between blocks
with strongest matrix elements being $0.0120$ from 
{\it Fellow} to {\it United States} and 
$0.0050$ from {\it Harvard University} to {\it University of Cambridge} (for both 
directions between blocks).
The link {\it Fellow} to {\it United States} is also the 
strongest indirect link in its off diagonal sub-block 
(for $\Gqr$) while the strongest direct link 
(for $\Grr$) is {\it Fellow} to {\it Harvard University}. Furthermore, the 
link {\it Harvard University} to {\it University of Cambridge} is also the 
strongest indirect link in its off diagonal sub-block 
(for $\Gqr$) while the strongest direct link 
(for $\Grr$) is {\it Harvard College} to {\it University of Cambridge}. 

Using the transition matrix elements of $G_{rr}+G_{qr}^{(nd)}$
we construct a network of effective friends shown in Figure~\ref{fig4}.
First, we select five initial nodes which are placed 
on a (large) circle: the two nodes with 
injection/absorption ({\it University of Cambridge} and {\it Harvard University}) 
and three other nodes with a rather top position 
in the $K_L$ ranking ({\it England}, (Town of) {\it Cambridge} and 
{\it United States}). For each of these five initial 
nodes we determine four friends 
by the criterion of largest matrix elements (in modulus) in the same 
column, i.e. corresponding to the four strongest links 
from the initial node to 
the potential friends. The friend nodes found in this way are added to the 
network and drawn on circles of medium size around their initial node 
(if they do not already belong 
to the initial set of 5 top nodes). The links from the initial nodes to their 
friends are drawn as thick black arrows. For each of the newly added nodes 
(level 1 friends) we continue to determine the 
four strongest friends (level 2 friends) which are drawn on small circles and 
added to the network (if there are not already present from a previous 
level). The corresponding links from level 1 friends to level 2 friends are 
drawn as thin red arrows. 

Each node is marked by the 
index $i$ from the first column of Table~\ref{table1}.
The colors of the nodes are essentially red for nodes with strong negative 
values of $P_1$ (corresponding to the index $i=1,\ldots,20$) and blue 
for nodes with strong positive values of $P_1$ (for $i=21,\ldots,40$). 
Only for three of the initial nodes we choose different colors which 
are olive for {\it US}, green for {\it England} and cyan for (the town of) 
{\it Cambridge}. 

The network of Figure~\ref{fig4} shows a quite clear separation of 
network nodes in two blocks associated to the two universities with a 
rather small number of links between the two blocks (e.g. US is a friend of 
England but not vice-versa).

\subsection{Case of  pathway Napoleon - Alexander I of Russia}

We illustrate the application of the LIRGOMAX and REGOMAX algorithms
on two other nodes of the Wikipedia network
with injection (pumping) at {\it Napoleon}
and absorption at {\it Alexander I of Russia}. 
The global PageRank indices of these two nodes are
$K=201$ (PageRank probability $P_0(201)= 0.0001188$)
and $K= 5822$ (PageRank probability $P_0(5822)= 1.389 \times 10^{-5}$ )
respectively. In contrast to the the previous example 
the two PageRank probabilities are rather different. However, this 
difference is compensated by our choice of the diagonal matrix $D$ with 
$D(201)=1/P_0(201)$ and $D(5822)=-1/P_0(5822)$ (other diagonal entries of $D$ 
begin zero). Again we 
determine the vector $V_0$ used for the computation of $P_1$
(see (\ref{eq_P1iterate})) by $V_0=G_0\,W_0$ 
where the vector $W_0=D\,P_0$ has the nonzero components $W_0(201) = 1$ and 
$W_0(5822) = -  1$. Furthermore, both $W_0$ and $V_0$ are orthogonal to the 
left leading eigenvector $E^T=(1,\ldots,1)$ of $G_0$. 

The top nodes of $P_1$, noted by index $i$, 
with 20 strongest negative and 20 strongest positive 
values are presented in Table~\ref{table2}. 
The ranking of nodes in decreasing order of $|P_1|$
given by the index $K_L$ is shown in the second column of Table~\ref{table2}. 
It is interesting to note that the injection node
{\it Napoleon} is only at position
$K_L=29$ with a significantly smaller
value of $|P_1|$ compared to 
{\it   Alexander I of Russia} at $K_L=3$,
{\it Russian Empire} at $K_L=1$ and
{\it  Saint Petersburg} at $K_L=2$.
But among the positive $P_1$ values {\it Napoleon} 
is still at the first position. 
We attribute this relatively small $|P_1|$ value of 
{\it Napoleon} compared to the nodes of the other block to significant
complex directed flows in the global Wikipedia network.
Also {\it Napoleon} has a significantly stronger
PageRank probability and thus this node produces a stronger
influence on {\it   Alexander I of Russia} than 
vice-versa.

In contrast to the previous case of universities 
Table~\ref{table2} contains mainly countries, a few towns and islands, 
and historical figures related in some manner to 
{\it Napoleon} or {\it   Alexander I of Russia}.

\begin{table}
\begin{center}
{\relsize{-2}
\caption{Same as in Table~\ref{table1} for English 
Wikipedia 2017 with injection (pumping) at {\it Napoleon} and absorption at 
{\it Alexander I of Russia}. }
\label{table2}
\begin{tabular}{rrrl}
\hline
$i$ & $K_L$ & $K$ &Node name \Tstrut\Bstrut\\
\hline\Tstrut
1 & 1 & 181 & Russian Empire \\
2 & 2 & 216 & Saint Petersburg \\
3 & 3 & 5822 & Alexander I of Russia \\
4 & 4 & 15753 & Paul I of Russia \\
5 & 5 & 3409 & Catherine the Great \\
6 & 6 & 158 & Moscow \\
7 & 7 & 17 & Russia \\
8 & 8 & 153 & Azerbaijan \\
9 & 9 & 9035 & Nicholas I of Russia \\
10 & 10 & 203707 & Elizabeth Alexeievna (Louise of Baden) \\
11 & 11 & 92 & Ottoman Empire \\
12 & 12 & 7 & Iran \\
13 & 13 & 177854 & Government reform of Alexander I \\
14 & 14 & 889 & Caucasus \\
15 & 15 & 31784 & Russo-Persian War (1804–13) \\
16 & 16 & 8213 & Alexander II of Russia \\
17 & 17 & 475 & Prussia \\
18 & 18 & 5764 & Dagestan \\
19 & 19 & 32966 & Serfdom in Russia \\
\Bstrut 20 & 20 & 131205 & Adam Jerzy Czartoryski \\
\hline\Tstrut
21 & 29 & 201 & Napoleon \\
22 & 52 & 192 & French Revolution \\
23 & 144 & 4 & France \\
24 & 149 & 12 & Italy \\
25 & 167 & 10611 & French Directory \\
26 & 180 & 24236 & Joséphine de Beauharnais \\
27 & 188 & 7361 & National Convention \\
28 & 189 & 21727 & French campaign in Egypt and Syria \\
29 & 195 & 2237 & Corsica \\
30 & 198 & 3166 & Louis XVI of France \\
31 & 199 & 7875 & Saint Helena \\
32 & 200 & 40542 & André Masséna \\
33 & 201 & 3353 & French Revolutionary Wars \\
34 & 203 & 11916 & Maximilien Robespierre \\
35 & 204 & 1241 & Louvre \\
36 & 205 & 69382 & Lucien Bonaparte \\
37 & 206 & 21754 & Coup of 18 Brumaire \\
38 & 207 & 15926 & French Republican Calendar \\
39 & 208 & 14931 & Jacobin \\
\Bstrut 40 & 209 & 7509 & Napoleonic Code \\
\hline
\end{tabular}
}
\end{center}
\end{table}

\begin{figure}
\begin{center}
\includegraphics[width=0.48\textwidth]{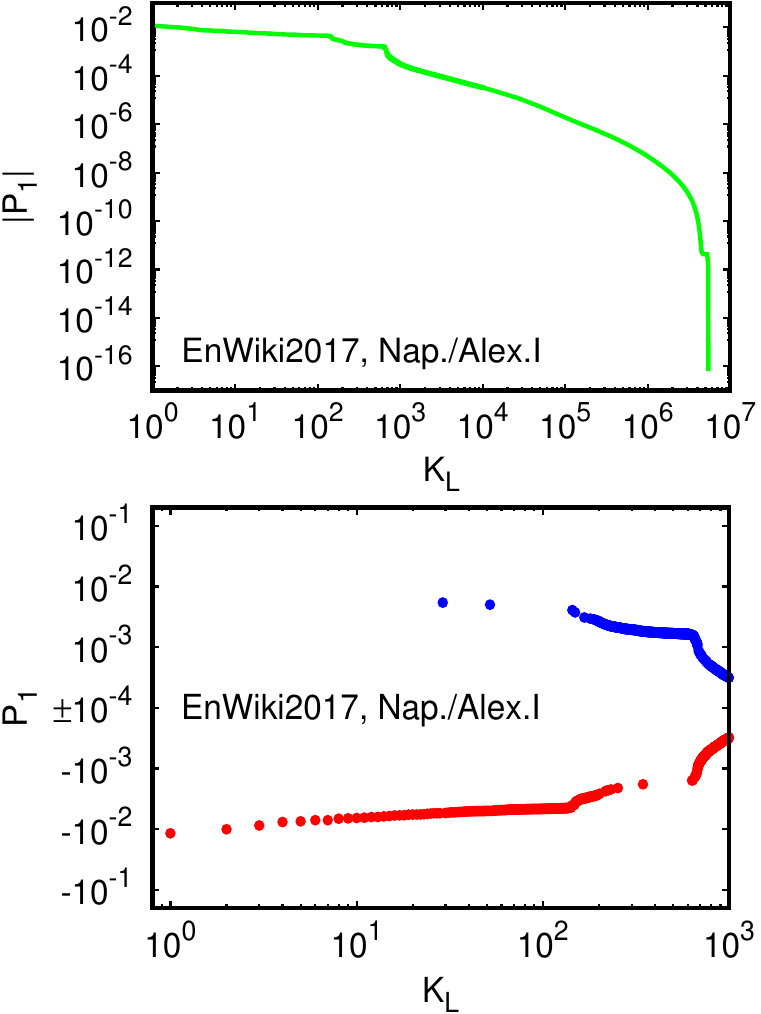}
\caption{Same as in Fig.~\ref{fig1} for the subgroup of 
Table~\ref{table2} corresponding to injection at {\it Napoleon} 
and absorption at {\it Alexander I of Russia}. }
\label{fig5}
\end{center}
\end{figure}

The dependence of the linear response vector $P_1$ on the index $K_L$ 
is shown in Figure~\ref{fig5}
(analogous to Figure~\ref{fig1}).
The decay of $|P_1|$ with $K_L$ is shown in the top panel,
being similar to the top panel of Figure~\ref{fig1}.
The values of $P_1$ with sign are shown in the bottom panel.
The difference of the $|P_1|$ values for  
{\it Napoleon} and {\it   Alexander I of Russia}
is not so significant but many nodes ($28$) from the block of 
{\it   Alexander I of Russia} have larger $|P_1|$ values than 
$|P_1|$ of {\it Napoleon}.

\begin{figure}
\begin{center}
\includegraphics[width=0.48\textwidth]{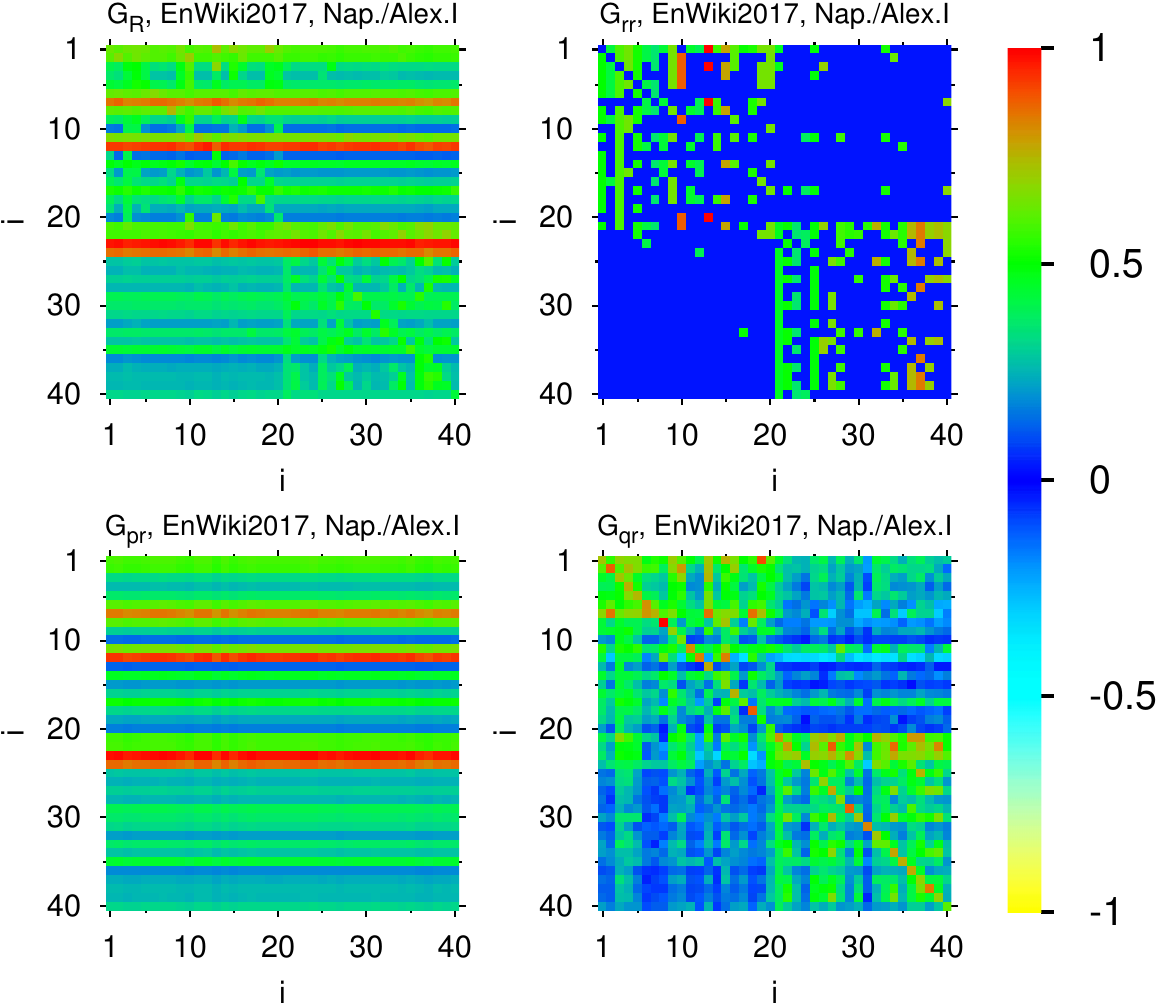}
\caption{As Fig.~\ref{fig2} for the subgroup of 
Table~\ref{table2} corresponding to injection at {\it Napoleon} 
and absorption at {\it Alexander I of Russia}. 
The relative weights of the different matrix components are 
$\Wpr=0.900$, $\Wrr=0.042$  and $\Wqr=0.058$.}
\label{fig6}
\end{center}
\end{figure}

The results for the reduced Google matrix of 40 nodes
of Table~\ref{table2} are shown in Figure~\ref{fig6}.
The strongest lines of transitions in $\GR$ and $\Gpr$
correspond to nodes with top PageRank positions
of the global Wikipedia network
being {\it France} at $K=4\ (i=23,\ K_L=144)$, {\it Iran} at 
$K=7\ (i=12,\ K_L=12)$, {\it Italy} at $K=12\ (i=24,\ K_L=149)$ 
and {\it Russia} at $K=17\ (i=7,\ K_L=7)$.
As explained above the structure of transitions
appears rather similar between $\GR$ and $\Gpr$. The weights
of all three components $\Gpr$, $\Grr$, $\Gqr$ are similar to those 
of the two universities (see caption of Figure~\ref{fig6}).

\begin{figure}
\begin{center}
\includegraphics[width=0.48\textwidth]{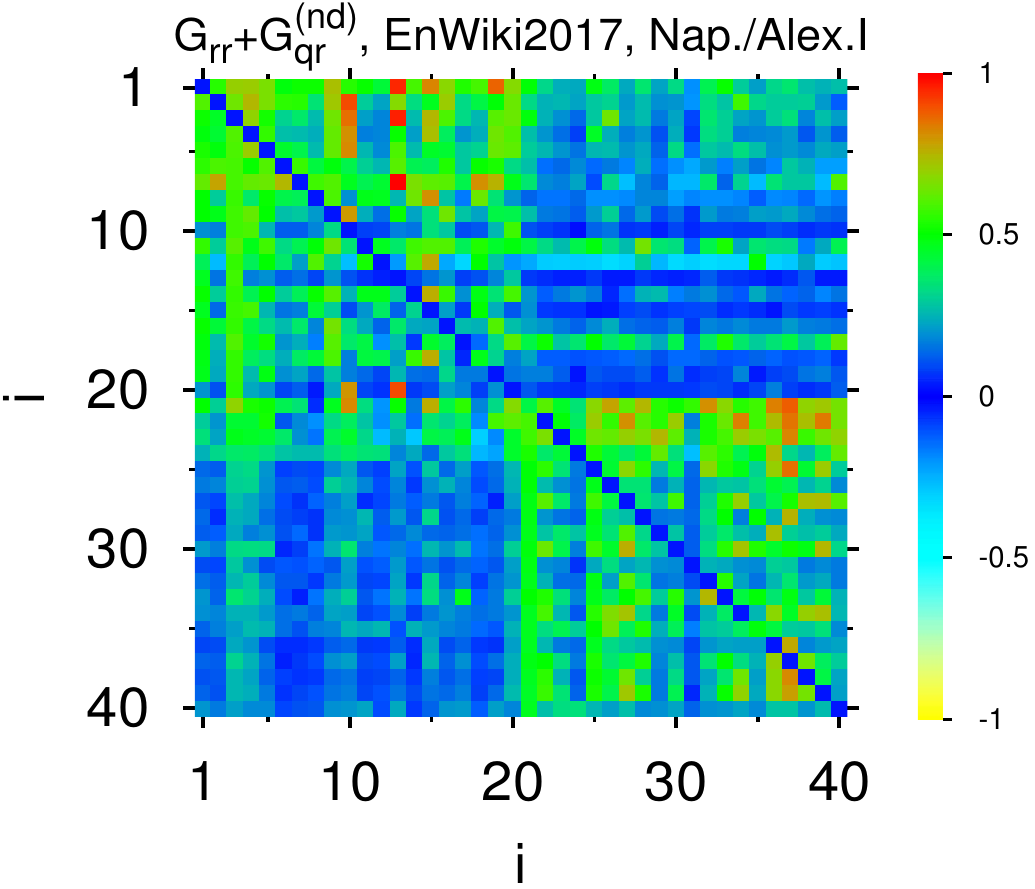}
\caption{Same as in Fig.~\ref{fig3} for the subgroup of 
Table~\ref{table2} corresponding to injection at {\it Napoleon} 
and absorption at {\it Alexander I of Russia}. 
The weight of $G_{rr}+G_{qr}^{(nd)}$ is $W_{rr+qrnd}=0.087$.}
\label{fig7}
\end{center}
\end{figure}

The components $\Grr$ and $\Gqr$, shown in Figure~\ref{fig6},
are also dominated by the two diagonal blocks related to the 
two initial nodes {\it Napoleon}
and {\it Alexander I of Russia}. There are only a few direct links
between the two blocks but the number of indirect links
is substantially increased.
The sum of these two components $G_{rr}+G_{qr}^{(nd)}$ 
is shown in Figure~\ref{fig7},
where the diagonal elements of $\Gqr$ are omitted.
The strongest couplings between the two blocks in
$G_{rr}+G_{qr}^{(nd)}$ are $0.009879$ for the 
link from {\it French campaign in Egypt and Syria} to {\it Ottoman Empire} 
and $0.02439$ for the link from {\it Elizabeth Alexeievna (Louise of Baden)} 
to {\it Napoleon} (for both directions between the diagonal blocks).

\begin{figure}
\begin{center}
\includegraphics[width=0.48\textwidth]{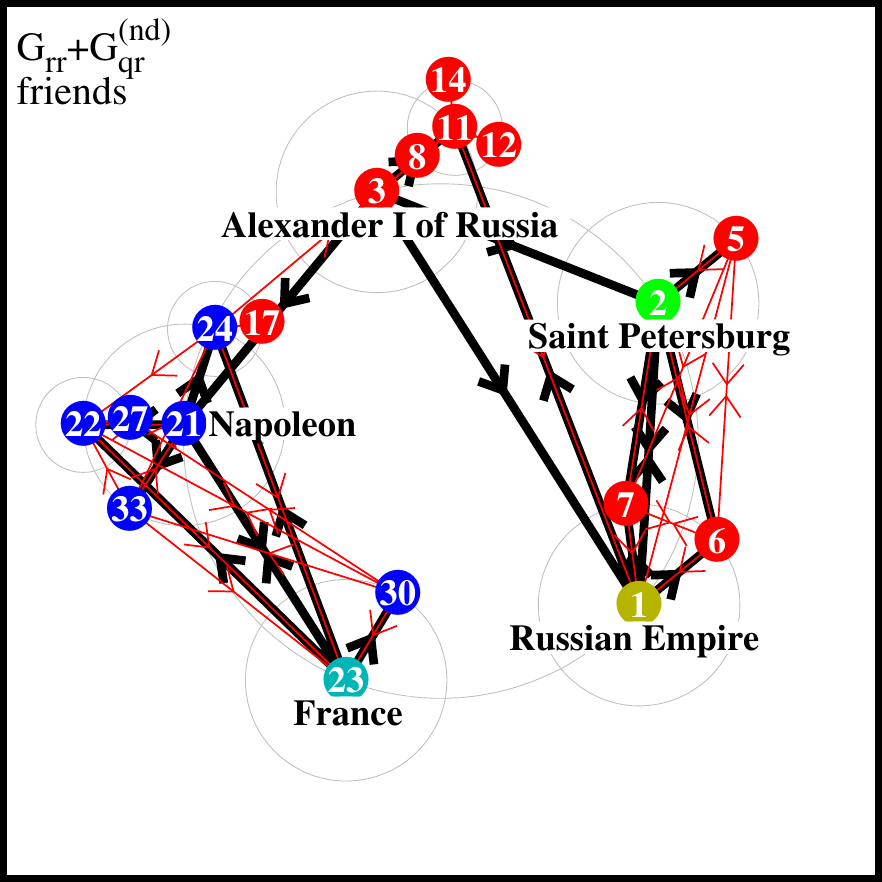}
\caption{Same as in Fig.~\ref{fig4} for the subgroup of 
Table~\ref{table2} corresponding to injection at {\it Napoleon} 
and absorption at {\it Alexander I of Russia}. }
\label{fig8}
\end{center}
\end{figure}

In analogy to Figure~\ref{fig4} we construct the network of friends
for the subset of Table~\ref{table2} shown in Figure~\ref{fig8}. 
As in  Figure~\ref{fig4}, we 
use the four strongest transition matrix elements of $G_{rr}+G_{qr}^{(nd)}$ 
per column to construct links from the five top nodes to level 1 friends 
(thick black arrows) and from level 1 to level 2 friends (thin red arrows).
As the five initial top nodes we choose 
{\it France} (cyan), {\it Russian Empire} (olive), {\it Saint Petersburg} (green), 
{\it Napoleon} (blue) and {\it Alexander I of Russia} (red); 
all other nodes of the {\it Napoleon}
block ($21 \leq i \leq 40$ in Table~\ref{table2}) are shown in blue,
and all other nodes of the {\it Alexander I of Russia}
block ($1 \leq i \leq 20$ in Table~\ref{table2})) are shown in red;
numbers inside the points correspond to the index $i$ of Table~\ref{table2}.

The network of Figure~\ref{fig8} also shows a clear two block structure 
with relatively rare links between the two blocks. The coupling between 
two blocks appears due to one link from  {\it Alexander I of Russia} to
{\it Prussia}, which even being red is more closely related
to the blued block of nodes. 

For both examples network figures constructed in the same way using the 
other matrix components 
$\GR$, $\Grr$ or $\Gqr$ (instead of $G_{rr}+G_{qr}^{(nd)}$) 
or using strongest matrix elements in rows (instead of columns) 
to determine follower networks are available at \cite{ourwebpage}. 

This type of friend/follower effective network schemes constructed from the 
reduced Google matrix (or one of its components) were already presented 
in \cite{politwiki} in the context of Wikipedia networks of politicians.

\section{Discussion}

We introduced here a linear response theory for a very generic model 
where either the Google matrix or the associated Markov process depends on 
a small parameter and we developed the LIRGOMAX algorithm to compute 
efficiently and accurately the linear response vector $P_1$ to the PageRank 
$P_0$ with respect to this parameter for large directed networks. 
As a particular application of this approach it is in particular possible 
to identify the most important and sensitive nodes
of the pathway connecting two initial groups of nodes 
(or simply a pair of nodes) with injection or absorption of probability. 
This group of most sensitive nodes can then be analyzed with the reduced 
Google matrix approach by the related REGOMAX algorithm which 
allows to determine effective indirect network interactions 
for this set of nodes. 
We illustrated the efficiency of the combined LIRGOMAX and REGOMAX 
algorithm for the English Wikipedia network of 2017 with 
two very interesting examples. In these examples, we use two initial nodes 
(articles) for injection/absorption, corresponding either to two 
important universities or to two related historical figures. As a result 
we obtain associated sets for most sensitive Wikipedia articles 
given in Tables~\ref{table1} and \ref{table2} with effective friend 
networks shown in Figures~\ref{fig4} and \ref{fig8}. 

As a further independent application the LIRGOMAX algorithm allows also 
to compute more accurately the Page\-Rank sensitivity with respect to 
variations of matrix elements of the (reduced) Google matrix as already studied in 
\cite{wrwu2017,wtn2019}. 

It is known that the linear response theory 
finds a variety of applications in statistical and mesoscopic physics
\cite{kubo,stone}, current density correlations \cite{kohn},
stochastic processes and dynamical chaotic systems \cite{hanggi,ruelle}.
The matrix properties and their concepts, like Random Matrix Theory (RMT),
find important applications for various physical systems 
(see e.g. \cite{guhr}). However, in physics 
one usually works with unitary or Hermitian matrices, like in RMT.
In contrast the Google matrices belong to 
another class of matricies rarely appearing in physical
systems but being very natural to the communication networks
developed by modern societies (WWW, Wikipedia, Twitter ...).
Thus we hope that the linear response theory for the Google matrix
developed here
will also find useful applications in the analysis of real directed networks. 

\section*{Acknowledgments}
This work was supported in part by the Programme Investissements
d'Avenir ANR-11-IDEX-0002-02, reference ANR-10-LABX-0037-NEXT 
(project THETRACOM);
it was granted access to the HPC resources of 
CALMIP (Toulouse) under the allocation 2019-P0110.

\end{document}